\documentclass[11pt,twoside,amstex,graphicx]{article}

\usepackage{asp2006}
\usepackage{epsf}
\usepackage{psfig}
\usepackage{lscape}
\usepackage{rotating}
\usepackage{amsmath}
\usepackage{graphicx,color}
\usepackage{subfigure}
\begin{document}
\title{Peculiar nulling in PSR J1738$-$2330}   
\author{Vishal Gajjar$^1$, Bhal Chandra Joshi$^1$, M. Kramer$^2$}   
\affil{$^1$National Centre for Radio Astrophysics(TIFR), Pune, India \\   
       $^2$Jodrell Back Centre for Astrophysics, University of Manchester, Manchester, UK}

\begin{abstract} 
Several pulsars show sudden cessation of pulsed emission, which is known
as nulling. The number of known nulling pulsars has not been significantly 
enhanced in the last decade, although the pulsar population has more 
than doubled following the Parkes multi-beam pulsar survey. A 
systematic follow-up study of the new pulsars, discovered in this 
survey, is being carried out by us at 325-MHz with GMRT. The 
peculiar nulling behaviour of PSR J1738$-$2330, observed as 
a part of this 325-MHz GMRT survey, is reported in this paper. 
The pulsar appears to show a periodic null-burst cycle with 
an upper limit to nulling fraction, of about 90 percent. 
The pulsed flux density declines by a factor 94 during the nulled 
pulses in this pulsar.
\end{abstract}


\section{Introduction}   

In some pulsars, the pulsed radio emission often abruptly stops for several 
periods. This phenomenon, called ``pulse nulling'', was discovered by 
Backer in 1970 (Backer 1970).
The fraction of pulses with no detectable emission 
is known as the nulling fraction (NF) and is a measure 
of degree of nulling in a pulsar. Several attempts have been made 
to correlate pulsar NF with various pulsar parameters, but till date no 
strong correlation exists. Ritchings (1976) concluded that NF in general 
increases with pulsar period, but pulsars close to the ``death line'' in 
$P-\dot{P}$ diagram are more likely to null. No such correlation between 
NF and age was reported in a given morphological class, when pulsars 
were grouped in different classes based on average profile morphology, 
implying stars that null more belong to profile classes, which are 
systematically older (Rankin 1986). A detailed study on 72 nulling pulsars 
suggested a correlation between NF and pulsar period (Biggs 1992). 
In recent past, more sensitive studies of nulling to estimate 
pulsar NFs more accurately have been carried out on a larger sample 
of pulsars (Vivekanand 1995; Wang et al. 2007) with this motivation. 
As the pulsar population has more than doubled after the recent Parkes 
multi-beam pulsar survey (Manchester et al. 2001), a 325-MHz 
survey of probable nulling pulsars, discovered in the Parkes multi-beam pulsar 
survey, with Giant Meterwave Radio Telescope (GMRT; Swarup et al. 1991) 
is being carried 
out by us to estimate their NFs. This paper presents the 
results on the peculiar nulling behaviour of PSR J1738$-$2330 observed 
in our survey.

The GMRT observations are described in Section 2 alongwith the 
analysis procedure used.
In Section 3, our results are presented and these are discussed in 
Section 4. 

\begin{figure}
 \centering
 \psfig{figure=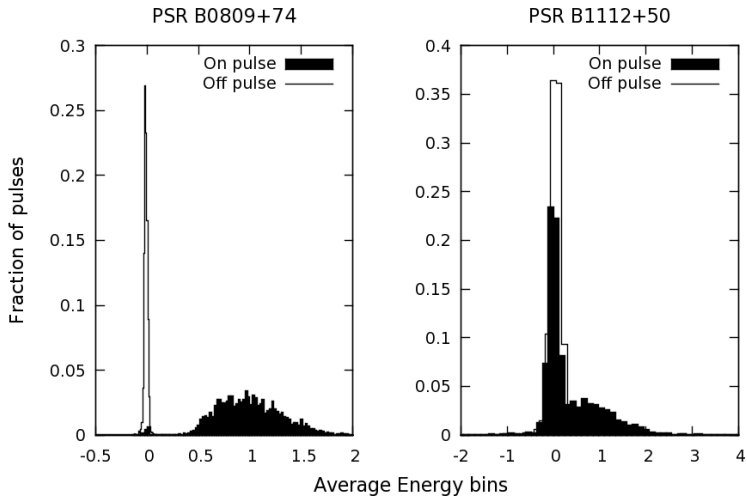,width=4in,angle=0}
 \caption{ON-pulse and OFF-pulse energy distributions  
          are shown in the left panel for PSR B0809+74 and 
          in the right panel for PSR B1112+50}
\label{fig1}
\end{figure}

\section{GMRT Observations}

For the initial phase of this survey, probable nulling pulsars were 
selected from those discovered in the Parkes multi-beam pulsar survey. 
A few well known nulling pulsars, such as PSR B0809+74 and PSR B1112+50, 
were also included as control pulsars. 
Observations of 20 sources, for about 2 hours each, were carried out 
with GMRT  
in a phased array mode at 325 MHz with 16 MHz of bandwidth. The two hands 
of circularly 
polarized voltages from 15 GMRT antennas including the compact central 
square and 3 arm antennas were added after compensating for phase delays 
forming a coherent sum. The sum of detected polarized powers was then 
recorded on a hard disk with a sampling time of 1 ms with 256 spectral 
channels across the band. The minimum detectable flux with a signal-to-noise 
ratio (SNR) of 6$\sigma$ was estimated to be around 7.4 mJy.  
So, these 
observations are more sensitive to single pulses compared to some of 
the previous observations (Wang et al. 2007).

\begin{figure}
 \centering
 \psfig{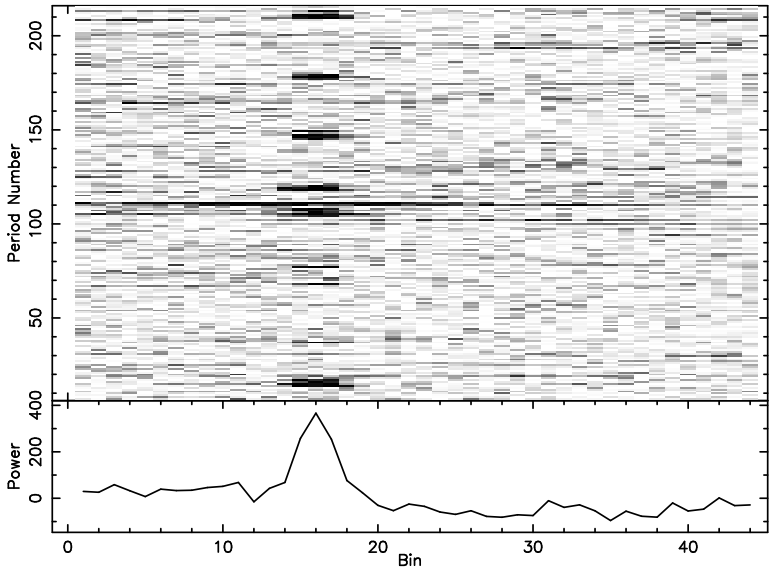}
\caption{The top plot shows the modulation of pulsar energy for 
successive 16 period subintegrations as a contour plot for 
bins 1 to 42 of the data dedispersed to 128 bins across the period. 
Bursts near subintegrations 15, 38, 70, 75, 105, 118, 145, 178 
and 210 are seen in the bins corresponding to the pulse window 
in the average profile, shown in the lower plot, interspersed 
with nulls.}
\label{fig3}
\end{figure}

The data for each pulsar were dedispersed using programs in a publically 
available package SIGPROC (http://sigproc.sourceforge.net). 
The resultant time series was folded 
every period to 128 phase bins across the pulse period. A baseline, 
estimated using bins 25 to 120 away from the pulse, was subtracted from 
the data for each period. Then, two sequences were formed by averaging the 
energies in bins 13 to 17 (ON-pulse energy) and bins 42 to 46 (OFF-pulse 
energy). The two sequences were then normalized by the mean pulse energy. 
The energy in the scaled sequences were also binned to 100 bins to form the 
ON-pulse and OFF-pulse energy distributions. An excess at zero energy  
in the ON-pulse energy distribution indicates the fraction of nulled 
pulses or NF of the pulsar. This can be estimated by removing 
a scaled version of OFF-pulse energy distribution at zero energy from 
the ON-pulse distribution. The procedure 
is similar to that used for detecting pulse nulling in single pulse 
sequences (See Ritchings 1976; Vivekanand 1995).

The procedure was first applied to data on two well known nulling pulsars, 
PSR B0809+74 and PSR B1112+50. The ON-pulse and OFF-pulse energy 
distributions for PSR B0809+74 are shown in the left panel of 
Figure \ref{fig1}. This 
prominently nulling pulsar has a clear bimodal ON-pulse energy 
distribution, with two peaks - one around the mean pulse energy and 
the other around zero pulse energy. The zero energy excess represents 
the nulled pulses and their ratio to total number of pulses 
gives an estimate for its NF. This was found to be  1 percent, 
consistent with previous studies (Lyne \& Ashworth 1983). In contrast, 
PSR B1112+50 is known to exhibit a large number of nulled pulses and 
this is evident from its ON-pulse and OFF-pulse energy distributions 
shown in the right panel of Figure \ref{fig1}. These distributions provide 
an estimate 
for this pulsar's  NF to be 61 percent, comparable to previously known 
results (Ritchings 1976).

\section{Nulling in PSR J1738$-$2330}

PSR J1738$-$2330 was discovered in the Parkes multi-beam pulsar survey 
(Manchester et al. 2001; Lorimer et al. 2006). It has a period of about 
1.9 s and a moderate DM (99.3 $cm^{-3}~pc$). It was observed on November 
22, 2008 for about two hours with GMRT. The dedispersed data were folded 
every 16  periods to 128 bins and these are shown in Figure \ref{fig3} 
alongwith its average 
profile. The pulsar seems to have periodic bursts, 
with an average duration of about 50 periods,  interspersed 
with nulls  of about 510 periods. Work is currently in progress to check 
this periodic behaviour using Fourier analysis. Recently, 
evidence for such periodic nulling has been reported 
in PSR B1133+16 (Herfindal \& Rankin 2007) and PSR J1752+2359 
(Lewandowski et al. 2004). If the periodic feature is 
confirmed, PSR J1738$-$2330 joins this 
class of pulsars.

\begin{figure}
 \centering
 \psfig{figure=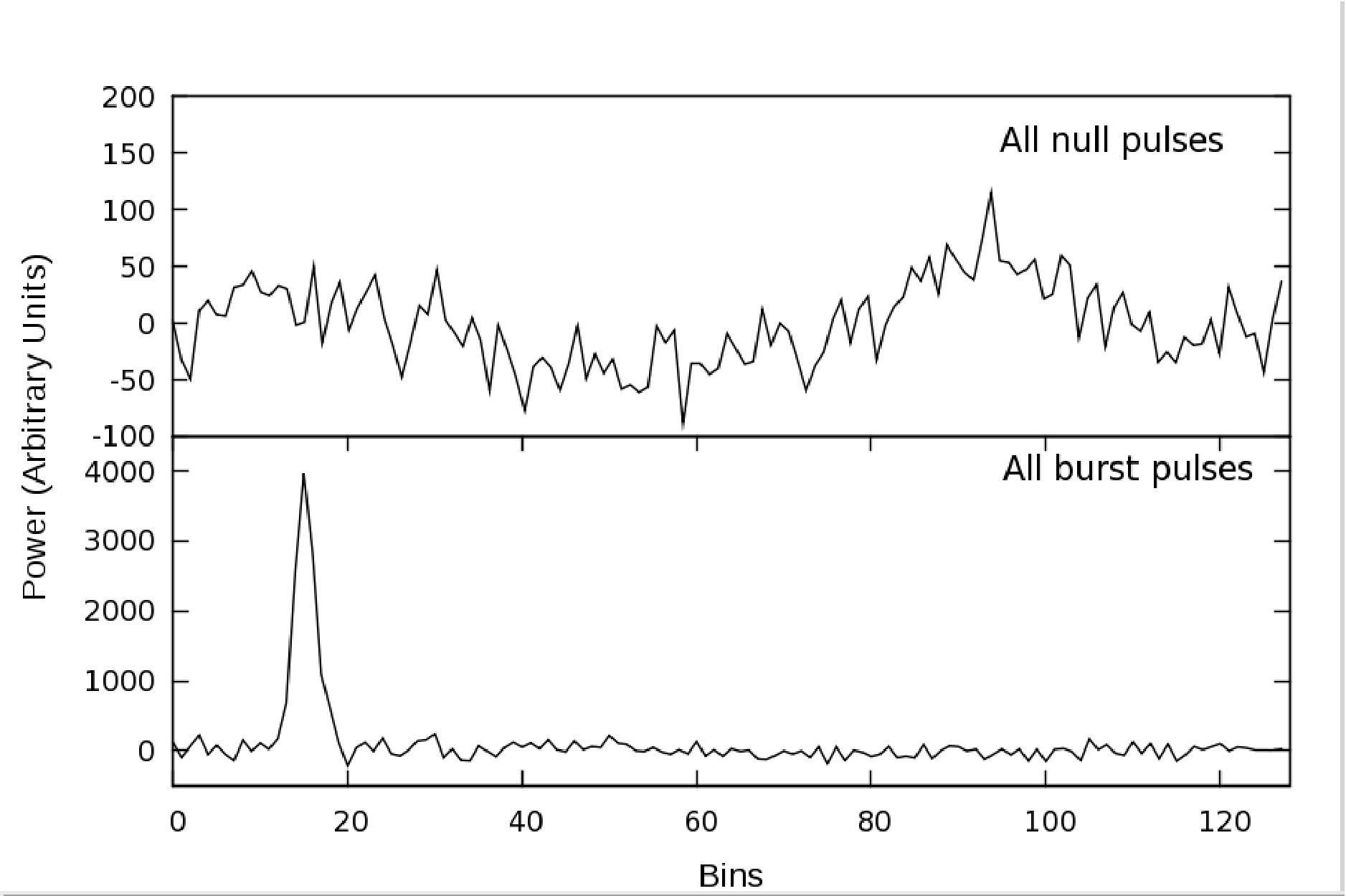,width=4in,angle=0}
\caption{Average profile of 3140 null pulses (top panel) and 
         293 burst pulses (bottom 
         panel) for PSR J1738$-$2330}
\label{fig4}
\end{figure}

It is also evident from Figure \ref{fig3} that the pulsar has a high 
NF. The null periods were visually identified from Figure \ref{fig3}. 
Average profiles of nulled pulses and burst pulses ({\it i.e.} periods
with detectable emission in the pulse window), culled from 
this single pulse analysis, were formed and are shown in Figure \ref{fig4}. 
It is clear from the average profile of all nulled pulses that 
there is no detectable weak emission during the pulse window. 
The average flux density in the pulsed emission during 
burst pulses is 94 times higher than that during nulled pulses, 
similar to results on other pulsars where this has been studied 
(Lyne \& Ashworth 1983; Vivekanand \& Joshi 1997).

\begin{figure}
 \centering
 \psfig{figure=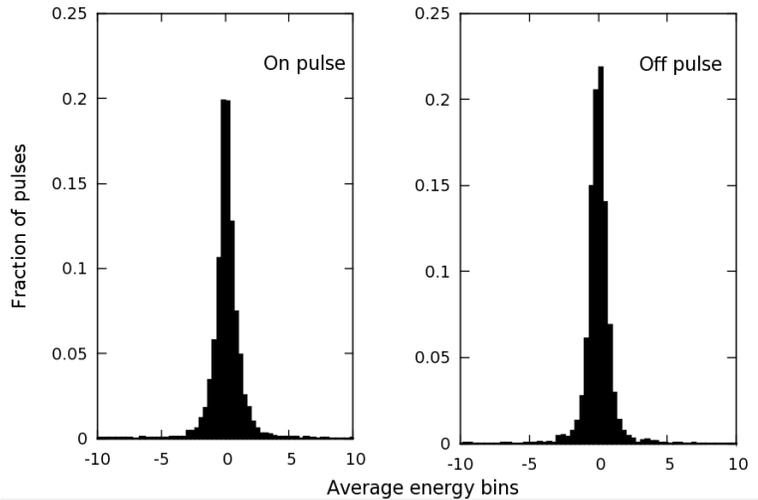,width=4in,angle=0}
 \caption{ON-pulse and OFF-pulse energy distributions for 
          PSR J1738$-$2330 }
\label{fig5}
\end{figure}

The ON-pulse and OFF-pulse energy distributions for this pulsar were 
obtained in a manner similar to PSR B0809+74 and B1112+50 and these 
are shown in Figure \ref{fig5}. These distributions are  
similar to PSR B1112+50, confirming a high NF for this pulsar. 
Our preliminary estimate for the upper limit to NF from this analysis 
is about 90 percent.

\section{Discussions and future work}

PSR J1738$-$2330 seems to show a periodic null-burst cycle. An upper limit 
to its NF of about 90 percent was obtained for the first time. 
The pulsed flux density declines by a factor of 94 during the nulled 
pulses in this pulsar.

Nulling is a poorly understood phenomenon. It could be due to 
a cessation of pair production in the polar gap (Ruderman \& Sutherland 
1975). In this framework, nulling behaviour of PSR J1738$-$2330 
suggests a periodic instability in the pair cascade in the 
polar gap. Another interesting possibility has been recently 
suggested by Herfindal and Rankin (2007), where a periodicity 
in nulling could be caused by a partially ignited sub-beam 
carousel. If this is indeed the case, a large number of sub-beams 
are not ignited in any carousel model proposed for this pulsar. 
Alternatively, the nulled pulses could be caused due to refraction 
effects or precession of the star. However, multi-frequency observations and 
polarization study of this pulsar is required to test these models 
and such studies are planned in future with GMRT.

\acknowledgements 
The Giant Meterwave Radio Telescope is a part of project by National Center for Radio Astrophysics which is funded by Tata Institute of Fundamental Research and Department of Atomic Energy.

\end{document}